\documentclass[]{emulateapj}
\usepackage{apjfonts}

\usepackage{epsfig, natbib, graphicx, color}

\input epsf

\newcommand{\Mpc}{\mbox{Mpc}}
\newcommand{\hMpc}{h^{-1}\mbox{Mpc}}
\newcommand{\msun}{M_\odot}
\newcommand{\boldm}[1]{\mathbf{#1}}

\newcommand{\avg}[1]{\left\langle #1 \right\rangle}
\newcommand{\Var}{\mbox{Var}}

\newcommand{\beq}{\begin{equation}}
\newcommand{\eeq}{\end{equation}}
\newcommand{\beqa}{\begin{eqnarray}}
\newcommand{\eeqa}{\end{eqnarray}}

\newcommand{\be}{\begin{equation}}
\newcommand{\ee}{\end{equation}}

\newcommand{\bea}{\begin{eqnarray}}
\newcommand{\eea}{\end{eqnarray}}

\newcommand{\psf}{{\mbox{psf}}}

\shortauthors{Rozo and Schmidt}
\shorttitle{Weak Lensing with Shear and Magnification}

\begin{document}
\title{Weak Lensing Mass Calibration With Shear and Magnification}
\author{Eduardo Rozo\altaffilmark{1,2} and Fabian Schmidt\altaffilmark{3}}
\altaffiltext{1}{Einstein Fellow, Department of Astronomy \& Astrophysics, The University of Chicago, Chicago, IL 60637}
\altaffiltext{2}{Kavli Institute for Cosmological Physics, Chicago, IL 60637}
\altaffiltext{3}{Theoretical Astrophysics, California Institute of Technology Mc 350-17, Pasadena, California 91125-0001, USA}

\begin{abstract}
We study how joint shear and magnification measurements improve the statistical precision of weak lensing mass calibration
experiments, relative to standard shear-only analysis.   For our magnification measurements,  we consider not only the impact
of lensing on the source density of galaxies, but also the apparent increase in source sizes.
The combination of all three lensing probes --- density, size, and shear --- can improve the statistical precision of mass estimates by
as much as $40\%-50\%$.  This number is insensitive to survey assumptions, though it depends on the degree of knowledge of the parameters 
controlling the magnification measurements, and on the value of the parameter $q$ that characterizes the response of the source population
to magnification.   Furthermore, the combination of magnification and shear allows for powerful 
cross-checks on residual systematics (such as point spread function corrections) at the few percent level.  

\end{abstract}

\section{Introduction}

Estimating cluster masses is a key component of many cosmological and astrophysical experiments.  Cosmologically, the abundance
of galaxy clusters as a function of their mass and redshift is widely recognized as a powerful probe of the fundamental physics behind
our accelerating universe \citep{detf06}, and currently provides one of the most robust estimates of the amplitude of the low redshift
matter power spectrum $\sigma_8$ \citep[e.g.][]{mantzetal08,henryetal09,vikhlininetal09b,rozoetal10a}.  Clusters have also been used to constrain
gravity on cosmological scales \citep{schmidtetal08,rapettietal,lombriseretal}.  
Astrophysically, comparison of
weak lensing mass estimates to those derived from X-ray or SZ observables can be used to study halo structure and 
the importance of non-thermal processes
in the intra-cluster medium 
\citep[e.g.][]{mahdavietal08,marroneetal09,umetsuetal09,morandietal10,meneghettietal10,molnaretal10}.   Moreover, precise cluster mass 
estimation significantly enhances the utility of galaxy clusters as a probe of galaxy evolution by allowing us to select galaxy populations
in halos of known masses.
The fact that all these applications require robust and well understood cluster mass estimates highlights the importance of cluster mass
estimation as a fundamental tool of contemporary astronomy.  

Today, weak gravitational lensing is widely identified as one of the most promising tools for robustly estimating cluster masses
\citep[for a review, see][]{bartelmannschneider01}. 
In the standard approach, one relies on the fact that gravitational lensing by foreground structures will shear background galaxies,
introducing a tangential pattern in the ellipticities of background galaxies about the center of the foreground overdensity.
These, however, are not the only observable signatures of weak gravitational lensing, as the latter also impacts the
number density, magnitudes, and sizes  of background galaxies.  In this letter, we explore how the combination of all three lensing measurements
can help estimate cluster masses more accurately and robustly  by breaking degeneracies between the desired physical quantities and systematic
nuisance parameters.   While galaxy sizes has already been proposed as a weak lensing signal \citep[e.g.][]{jain02}, and 
joint shear and density analysis has already been explored in the literature --- both in the context of cosmic shear \citep{zhangetal10} and for weak lensing
mass calibration \citep[for some examples see][]{Broadhurst1994,schneideretal00,vanwaerbekeetal,umetsuetal10} ---
we consider for the first time all these observational signatures simulteneously.  Moreover, we do not ignore the covariance between the various
observational signatures of gravitational lensing, as has been done in previous studies.

As we were finishing this letter, we learned of an independent analysis that is very much on the same spirit as ours, but which focuses on cosmic
shear, and in particular on the ability of joint lensing probes to self-calibrate multiplicative shear errors \citep{vallinottoetal10}.
The two analysis are qualitatively similar and reach very similar conclusions, namely that joint lensing analysis
are  a promising avenue of self-calibrating systematic uncertainties in weak lensing experiments.


\section{The Weak Lensing Observables}

We consider the problem of weak lensing mass calibration using three different observables.  
All of our estimators use radial bins $r_\alpha$, and global normalization.  We
adopt the notation of \cite{paperA}.  
The first observable is $n_\alpha$, the mean source density in an annulus: 
\be
n_\alpha =  \frac{1}{A_\alpha} \sum_i \Delta\Omega n_i W^\alpha_i.
\ee
In the above expression, 
we have pixelized all space, the index $i$ runs over all spatial pixels, $n_i$ is the source density at pixel $i$ (so that $n_i\Delta\Omega$ 
is either 0 or 1), $W^\alpha_i$ is the top hat filter defining the annulus of interest, and $A_\alpha$ is the area of said annulus.  

The second observable we consider is the area-averaged
number weighted shear of galaxies in an annulus, which we denote $h$.  Specifically, we write
\be
h_\alpha =  \frac{1}{A_\alpha}\sum_i \Delta \Omega n_i e_i W^\alpha_i
\ee
where $e_i$ is the galaxy ellipticity field at pixel $i$. 

Finally, the third observable is the number-weighted average apparent area of source galaxies $s_\alpha$, defined
via
\be
s_\alpha = \frac{1}{A_\alpha} \sum_i \Delta\Omega n_i a_i W^\alpha_i
\ee
where $a_i$ is the area of the source galaxy under consideration.  Note that one can choose among different 
possible estimators to capture magnification via galaxy sizes (\cite{jain02}).  Generally, the different
estimators have similar statistical and systematic properties, and our conclusions do not depend
on the precise choice of estimator.  

For simplicity, we assume sources are unclustered so that
\be
\avg{n_i n_j} = \bar n^2 \mu_i^{q/2}\mu_j^{q/2}\left(1+\delta_{ij}\frac{1}{\bar n \mu_i^{q/2} \Delta\Omega} \right)
\ee
where $q$ is the parameter governing the impact of lensing bias, i.e. $n_{obs} = \mu^{q/2}n_{true}$
\citep{schmidtetal09}, and $\mu_i$ is the magnification in the direction of galaxy $i$.  
Source clustering will be briefly discussed in section \ref{sec:conclusions}.
We further assume the ellipticities of the sources satisfy 
\bea
\avg{e_i} & = & g_i \\
\avg{e_ie_j} & = & g_ig_j+\delta_{ij}\sigma_e^2/2.
\eea
As for the source area, it is obviously of critical importance to consider the impact of the Point Spread Function (PSF) on 
the apparent galaxy area.  For Gaussian sources and PSF, the PSF adds to the apparent size, so we assume
\be
a_{obs} = \mu a_{true} + a_\psf
\ee
where $a_\psf$ characterizes PSF broadening and where we have included the impact of 
magnification on the apparent area of the source.   Usually, size estimates are corrected
(deconvolved) for PSF effects, so in principle, the PSF offset $a_\psf$ is identically zero.  In practice,
imperfect PSF deconvolution can lead to systematics offsets, so we will retain $a_\psf$
as a systematics parameter.

Defining
\bea
\bar a & = & \avg{a_{true}} \\
\sigma_a^2 & = & \Var(a_{true})/\bar a^2 \\
\epsilon & = & a_\psf/\bar a 
\eea
one has that
\bea
\avg{a_i} & = & \bar a (\mu+\epsilon) \\
\avg{a_ia_j} & = & \avg{a_i}\avg{a_j} + \delta_{ij} \mu^2 \bar a^2 \sigma_a^2
\eea
with $\epsilon$ characterizing a possible PSF correction bias.

With these assumptions, the expectation values and correlation matrix of our observables are given by
\bea
\avg{n_\alpha} & = & \bar n \mu_\alpha^{q/2} \\
\avg{h_\alpha} & = & \bar n \mu_\alpha^{q/2}g_\alpha \\
\avg{s_\alpha} & = & \bar n \bar a \mu_\alpha^{q/2}(\mu+\epsilon),
\eea
\be
C_{\alpha\beta}^{ab} = \frac{\delta_{\alpha\beta}\: \bar n}{A_\alpha} \mu_\alpha^{q/2} \left( \begin{array}{ccc}
 1 & g_\alpha           & \bar a (\mu_\alpha +\epsilon)   \\
 \sim  & g_\alpha^2 + \frac{\sigma_e^2}{2}  & \bar a (\mu_\alpha +\epsilon) g_\alpha \\
 \sim  & \sim                                   & \bar a^2 \left[ (\mu_\alpha+\epsilon)^2+\mu_\alpha^2 \sigma_a^2 \right]
\end{array}\right)
\ee
where $a,b \in \{ n,h,s \}$, and the symbol $\sim$ signifies that the covariance matrix is symmetric.
In deriving the above equations we have made use of the narrow-bin approximation, that is, we have assumed all lensing quantities are constant across the radial
bins employed for the measurements.

It is useful to allow for the possibility of the source density of galaxies used for the density measurement $n$ to be different from the source
density used for the shear and area measurements.   In such a scenario, the source galaxies employed for the density measurement 
that are not included in the source density measurement will in general have a different value for the $q$ parameter.  
Assuming the relative densities are related by a factor $\lambda$ such that the source density for measuring $n$
is $\lambda \bar n$, and letting $Q$ denote the lensing parameter of the sources employed in the density measurement,
the above equations are modified only via
\bea
\avg{n_\alpha} & = & \lambda \bar n \mu_\alpha^{Q/2} \\
C_{\alpha\beta}^{nn} & = & \delta_{\alpha\beta} \frac{\lambda \bar n}{A_\alpha} \mu_\alpha^{Q/2} \\
C_{\alpha\beta}^{nh} & = & \delta_{\alpha\beta} \frac{\bar n}{A_\alpha} \mu_\alpha^{q/2}g_\alpha \\
C_{\alpha\beta}^{na} & = & \delta_{\alpha\beta}  \frac{\bar n}{A_\alpha} \bar a\mu_\alpha^{q/2} (\mu+\epsilon)
\eea
Note that the covariance between density and other observables goes as $\mu^{q/2}$ since only galaxies that are shared
by the two measurements are employed, so the appropriate lensing parameter is $q$ rather than $Q$.

We will be primarily interested on how a joint analysis of the above lensing observables compares 
to the standard weak lensing shear analysis, which uses as its observable the mean galaxy
ellipticity in annuli,
\be
\bar e_{\alpha} = \frac{\sum_i \Delta\Omega n_i e_i W^\alpha_i}{\sum_i \Delta\Omega n_i W^\alpha_i}.
\ee
Note that this estimator uses a location-dependent normalization, rather than a global normalization.  
The mean and variance of this observable are (see \cite{paperA})
\bea
\avg{\bar e_{\alpha} } & = & g_\alpha \\
C^{ee}_{\alpha\beta} & = & \delta_{\alpha\beta} \frac{1}{\bar n A_\alpha} \frac{\sigma_e^2}{2}.
\eea
We use the above set of equations to set the baseline against which the results from a joint analysis are to be compared against.


\subsection{Signal-to-Noise Considerations}
\label{sec:s2n}

Before we forecast the precision with which weak lensing observables can constrain cluster masses,
it is worth considering the signal-to-noise of each of our three observables: density ($n$), shear ($h$),
and size ($s$).  For simplicity, we will work in the weak lensing limit 
and assume negligible systematics ($\epsilon=0$).  We find
\bea
(S/N)_n & = & (\bar n A)^{1/2} \lambda^{1/2} Q\kappa \\
(S/N)_h & = & (\bar n A)^{1/2} \sqrt{2} \frac{\gamma}{\sigma_e} \\
(S/N)_s & = & (\bar n A)^{1/2} \frac{q+2}{[1+\sigma_a^2]^{1/2}} \kappa.
\eea

There are two key points to take away from the above results:
\begin{enumerate}
\item The signal-to-noise of all measurements scales as $(\bar n A)^{1/2}$.  Consequently, the {\it ratio} of the mass uncertainty between any two
experiments is independent of the assumed source density.  We use this to our advantage by normalizing all of our uncertainties relative 
to that of a standard
shear-only experiments. Note, however, that when including priors
our quantitative results are not entirely $\bar n$-independent since the relative importance of priors varies with $\bar n$.  That said, we find 
this $\bar n$ dependence on the relative errors to be rather modest for moderate changes in $\bar n$.
\item The signal-to-noise of the size measurement is independent of $\bar a$.  Consequently, all of our results are fully independent of 
the fiducial value of $\bar a$, a fact which we have explicitly checked.
\end{enumerate}

One last important result that can be garnered from the above signal-to-noise estimates is that large boost parameters $\lambda$
are not necessarily optimal for density measurements.  For instance, in current experiments,
if source selection occurs close to the confusion limit then $\lambda$
can be as large as $\lambda\approx 4$, but the
$Q$ parameter tends to be relatively small, $Q\approx 0.5$.  On the other hand, if one requires that only sources well 
above the confusion limit be included in the source population, as is done for shear and area measurements, then $\lambda$
is modest, with $1 \leq \lambda\lesssim 1.5$, but
$Q$ can be significantly larger, with $Q\approx q \approx 1.5$ \citep{schmidtetal09}.  
Comparing the former choice with $(Q,\lambda)=(0.5,4)$ to the latter case of $(1.5,1.5)$, we find
that the second choice is expected to have nearly twice the signal-to-noise.  Consequently, from now on we will make
the simplifying approximation that $Q=q$ and the boost factor $\lambda$ is moderate, no more than $1.5$.
In practice, given an empirically determined relation $Q(\lambda)$, one may find the source cuts
which maximize the signal-to-noise of the density measurement by setting
\be
\frac{d(Q\sqrt{\lambda})}{d\lambda} = 0.
\ee
In principle, if $Q$ keeps increasing quickly with decreasing $\lambda$, it is possible that the density measurement is optimized using fewer source
galaxies than those employed in the shear measurement.


\section{Fiducial Model and Fisher Matrix Forecast}

We assume halo mass profiles are described by the standard \citet[NFW,][]{nfw96} form, which is determined by two
parameters, the halo mass $M_{200m}$, defined relative to the mean matter density, and its concentration. 
We set $M_{200m}=10^{15}\ \msun$ and $c=5$, and adopt lens and source redshifts
of $z_L=0.3$ and $z_s=1.0$ respectively.  For our background cosmology we assume flat $\Lambda$CDM model
with $\Omega_m=0.28$ and $h=0.7$.  Distances are in physical units in $\Mpc$, {\it not} $\hMpc$.

We assume each of our three observables --- $n$, $h$, and $s$ ---  is measured in 
narrow, logarithmically spaced radial bins of width $\Delta \ln R = 0.05$, extending from 
$R_{min}=0.2\ \Mpc$ to $R_{max}=2\ \Mpc$.  Our results are largely insensitive to $\Delta \ln R$, and only mildly sensitive
to the innermost radius $R_{min}$ so long as the strong lensing region is excluded.  
There is some sensitivity to the assumed maximum radius $R_{max}$, in the sense that larger $R_{max}$ values
decreases the relative importance of density and size measurements.  


\begin{deluxetable}{c c l}
\tablewidth{0pt}
\tablecaption{\label{tab:pars} Fisher Matrix Parameters}
\tablehead{Parameter & Fiducial Value & Explanation}
\startdata
$M$ & $10^{15}\ \msun$  & Halo mass\\
$c$ & $5$ & Halo concentration \\
$\bar n$ & $10\ \mbox{sources}/\mbox{arcmin}^2$ & Mean source density \\
$q$ & $1.5$ & $h$ and $s$ lensing parameter\\
$\bar a$ & Irrelevant & Mean source size\\
$\epsilon$ & $0.0$ & Systematic Size PSF Residual \\
\hline
\hline
$\sigma_e$ & $0.4$ & Std. Dev. of source ellipticity \\
$\sigma_a$ & $1.2$ & (Std. Dev. of source area)/$\bar a$ \\
$\lambda$ & $1.5$ & Density measurement source boost \\
$z_L$ & $0.3$ & Halo redshift \\
$z_s$ & $1.0$ & Source redshift 
\enddata
\tablecomments{
Quantities below the horizontal lines are held fixed and assumed to be known a priori
throughout the entire manuscript.
}
\end{deluxetable}


For the lensing parameters, we set $q=1.5$ and $\epsilon=0$.   We also adopt $\sigma_e=0.4$ and $\sigma_a=1.2$ for
the intrinsic ellipticity and size dispersion.  The former value is typical of ground-based shear experiments, while the latter is estimated
from an ongoing study of size-lensing with data from the COSMOS survey \citep{cosmos}.  
In accordance with the discussion in section \ref{sec:s2n}, we force
the parameter $Q$ to be identical to $q$ --- i.e. we assume the same size and magnitude cuts in source selection for 
all measurements --- and we set $\lambda=1.5$.  For the source density,
we assume $\bar n=10\ \mbox{galaxies}/\mbox{arcmin}^2$, as expected for the DES, though we note our results are fairly
insensitive to the precise value of $\bar n$.  The fiducial value for $\bar a$ is irrelevant as our results are $\bar a$-independent
(see section \ref{sec:s2n}).  
That said, the precision with which $\bar a$ is known is important, so we do
consider how uncertainties in $\bar a$ impact our results.
 
The Fisher matrix for a joint density+shear+size weak lensing experiment is given by
\be
F_{ij} = \sum_\alpha \sum_{a,b} (C^{a,b})^{-1}_{\alpha,\alpha} \frac{\partial \avg{a_\alpha}}{\partial p_i}\frac{\avg{b_\alpha}}{\partial p_j}
\ee
where $a,b$ loop over our three observables, and $\alpha$ indexes the radial bins employed in the experiment.
The vector $\boldm{p}$ is the vector of parameters of interest, which at minimum includes halo mass and concentration, but can
also include additional nuisance parameters such as the mean source density $\bar n$ or the lensing bias parameter $q$.
The corresponding covariance matrix is given by $C=F^{-1}$. 
Table \ref{tab:pars} summarizes the parameters considered in our Fisher Matrix analysis.


\section{Results}
\label{sec:results}


\begin{figure}
\epsscale{1.20}
\plotone{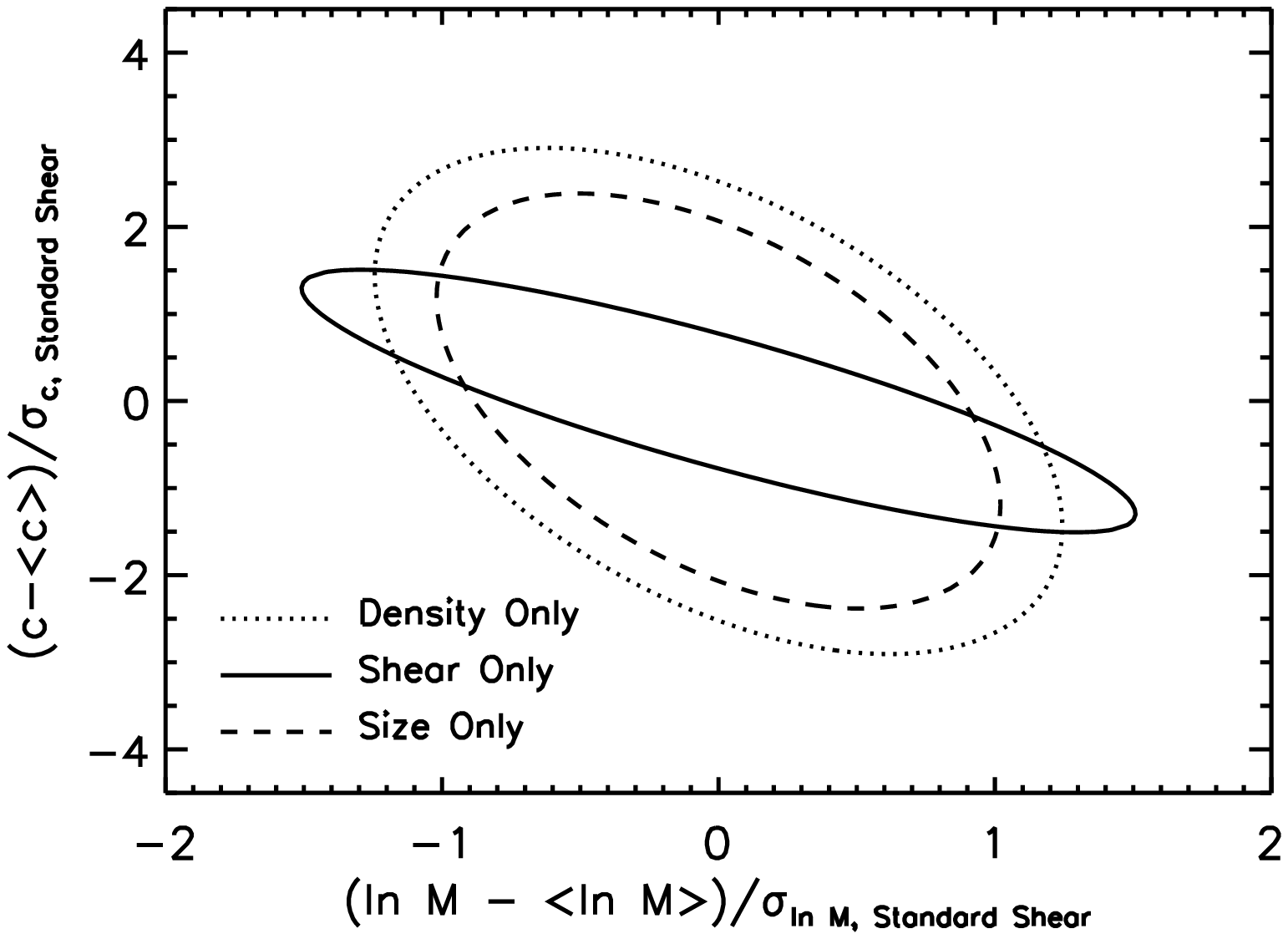}
\plotone{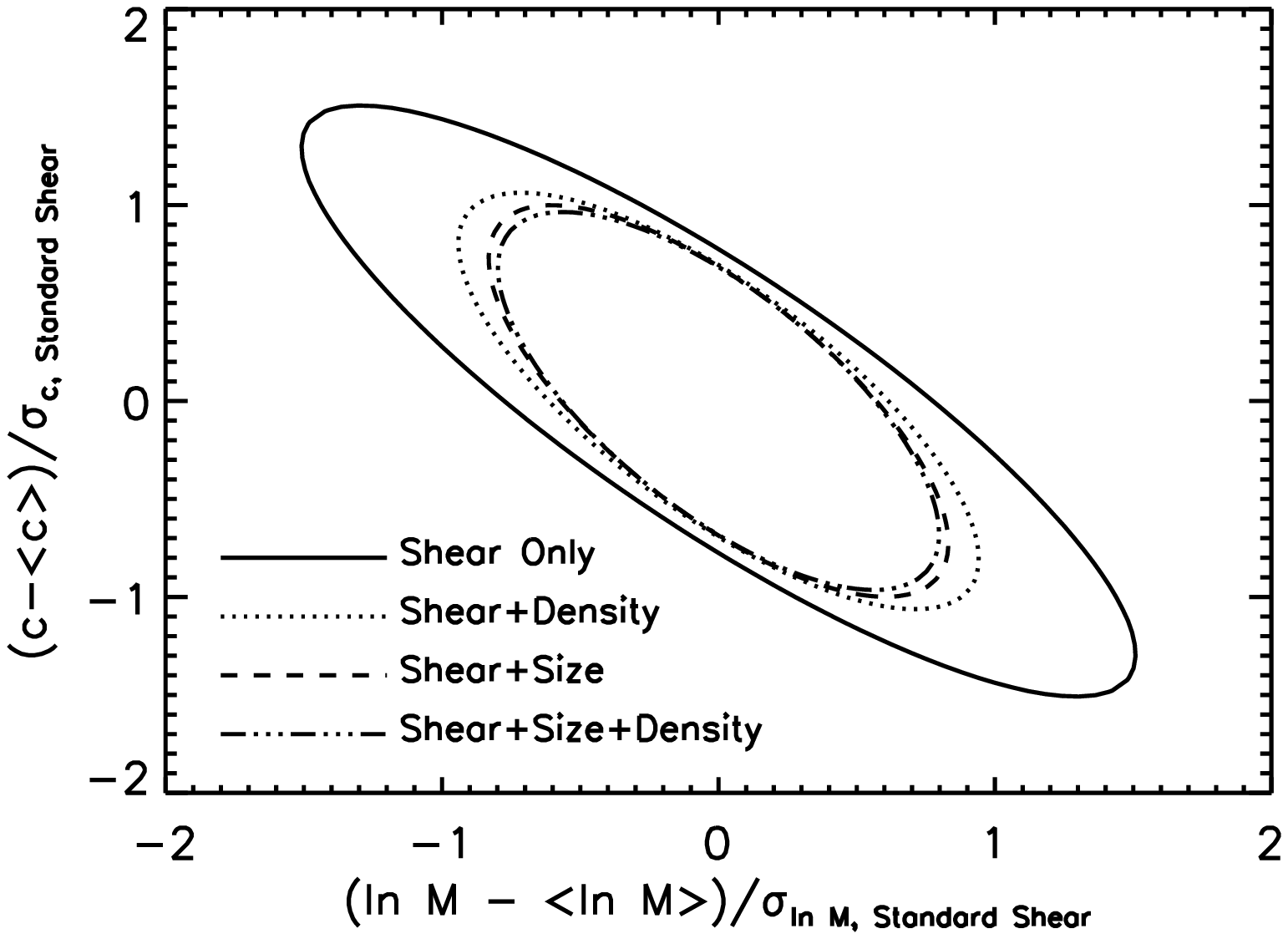}
\caption{$68\%$ error ellipses in the halo mass--concentration plane for a variety of weak lensing experiments as labelled.
The top panel treats each observable independently while the bottom panel considers a joint analysis.  
The axes are chosen such that the input halo mass and concentration occur at the origin, and the scale of the axis
is in units of the marginalized $1\sigma$ error for a standard shear-only analysis.   
All parameters other than halo mass and concentration are held fixed.  Joint analyses have the potential to increase the
precision of weak lensing mass calibration experiments by $\approx 35\%$.
}
\label{fig:errors}
\end{figure} 


Figure \ref{fig:errors} shows our forecasted $68\%$ error ellipse in the mass--concentration plane for a variety of weak lensing
mass calibration experiments.  
The mass and concentration axes have been displaced so that the input parameters
occur at the origin, and the axis are in units of the marginalized standard deviation of each parameter for a standard shear
analysis.  When normalized
in this fashion, our results are insensitive to the source density $\bar n$.  All parameters other than halo mass and concentration
are held fixed (section \ref{sec:priors} will relax this assumption).

The top panel in Figure \ref{fig:errors} considers each of our observables as an independent weak lensing experiment.  The solid line
corresponds to the standard shear-only analysis, and is nearly identical to that obtained using the observable $h$ alone (not shown).  
The dotted and dashed
lines correspond to density ($n$) and size ($s$) respectively.  We find shear measurements lead to the tightest error ellipses,
but size and density measurements can achieve comparable and even higher precision than the shear measurement with respect
to mass estimation.  This conclusion does depend on our fiducial assumptions, and a lower $q$ value can render shear the most 
precise estimator.  The important point to emphasize though is that broadly speaking the various estimators have comparable
precision.

The bottom panel in Figure \ref{fig:errors} illustrates how the precision of a shear experiment improves as we include 
additional observables.  
The curves shown are for a shear only (solid), shear+density (dotted), shear+size (dashed),
and shear+size+density experiment (dash-triple dot).  We find that a joint analysis of all three observables significantly improves
the precision of weak lensing mass calibration experiments. Specifically, the uncertainty in the best fit log-mass is reduced
by almost $50\%$, with $\sigma_{\ln M, \mbox{\small joint}} / \sigma_{\ln M, \mbox{\small shear only}} = 0.53$.  
From now on, whenever we quote mass uncertainties, they will always be normalized in this way, so that $\sigma_{\ln M}=1$
mean that the analysis under consideration achieves the same precision as a shear-only analysis.


\subsection{Priors and Systematics Self-Calibration}
\label{sec:priors}

We have seen that joint shear+density+size analysis are in principle significantly more powerful than standard shear-only analysis.
In practice, however, the various lensing parameters that impact the density and shear signals are not perfectly known a priori.
We now explore to what extent our results generalize when faced with uncertainties in $\bar n$, $q$, $\bar a$, and $\epsilon$.


\begin{figure}
\epsscale{1.2}
\plotone{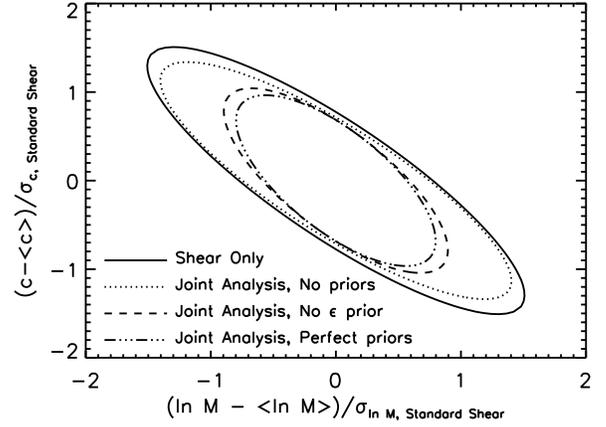}
\caption{$68\%$ confidence contours for a joint shear+density+size weak lensing mass calibration experiment.
The origin corresponds to the input halo mass and concentration, and the scale of the axes
is in units of the marginalized $1\sigma$ error for a standard shear-only analysis.  The solid ellipse is the
standard shear-only results, while the remaining ellipses are for a joint analysis with various priors.  They are 
as follows.  Dashed: fixed $\bar n$, $q$, and $\bar a$, with no prior on $\epsilon$, resulting in $\sigma_{\ln M}=0.59$.
Dotted: no priors on $\bar n$, $q$, $\bar a$, or $\epsilon$, resulting in $\sigma_{\ln M}=0.93$.  
Dash--triple-dot: fixed $\bar n$, $q$, $\bar a$, and $\epsilon$, resulting in $\sigma_{\ln M}=0.53$.
}
\label{fig:err_wpriors}
\end{figure} 


Figure \ref{fig:err_wpriors} compares the $68\%$ confidence contours of a standard shear-only analysis (solid)
to those of a joint analysis with a variety of priors for the nuisance parameters: perfect knowledge (dash--triple-dot),
fixed $\bar n$, $q$, and $\bar a$ but no prior on $\epsilon$ (dashed), and no priors on any lensing parameter
(dotted).  Even in this last case, we find some small improvement in the precision of a weak lensing mass calibration
experiment, with $\sigma_{\ln M}=0.93$, i.e. a $7\%$ improvement over the shear-only case.

We want to determine now the level of precision required on a priori knowledge of our nuisance parameters
in order to saturate the perfect knowledge limit illustrated in Figure \ref{fig:err_wpriors}.  Since the density
measurement is affected by only two of our four nuisance parameters, we begin by focusing on a shear+density
experiment.


\begin{figure}
\epsscale{1.2}
\plotone{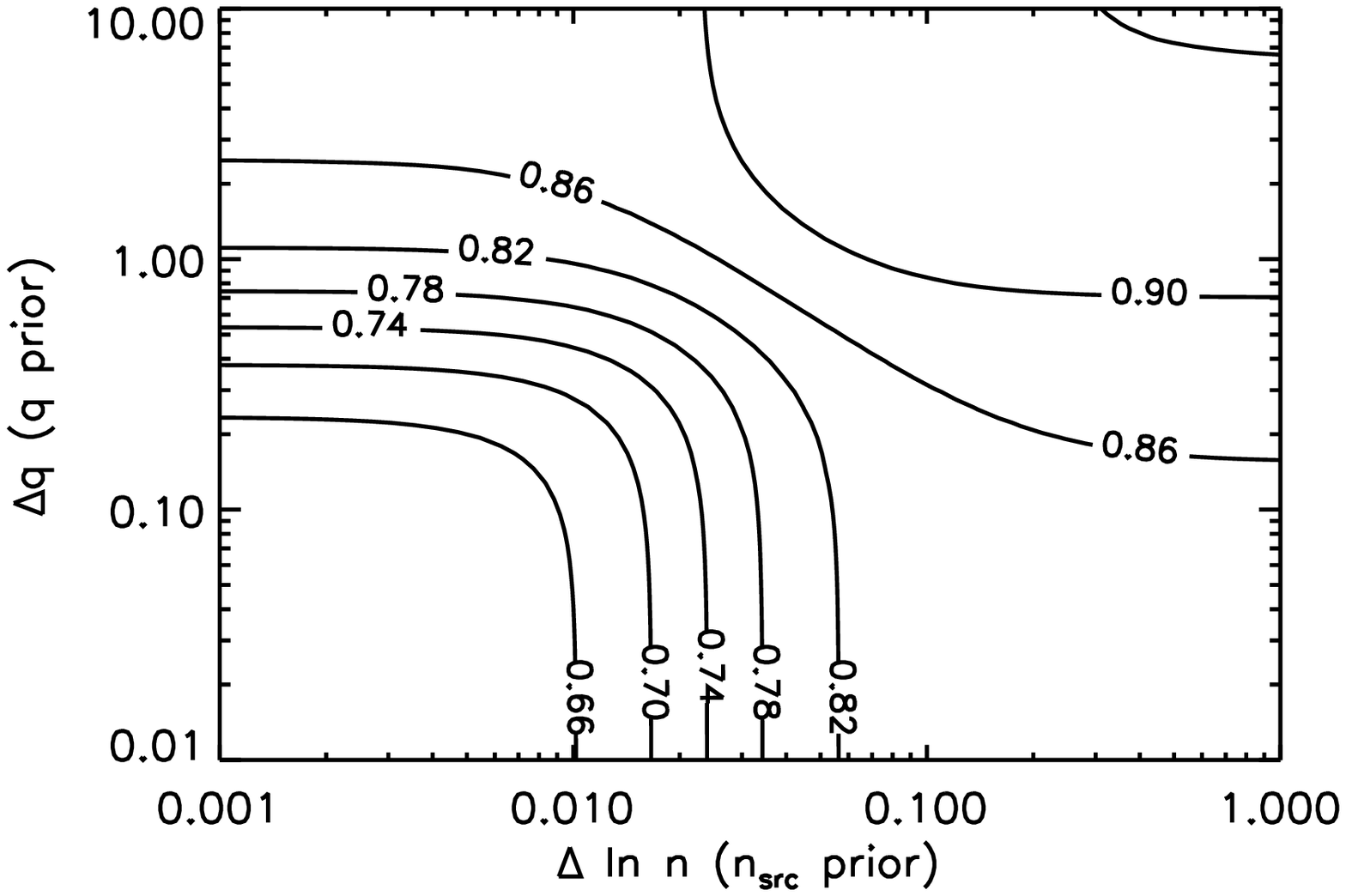}
\plotone{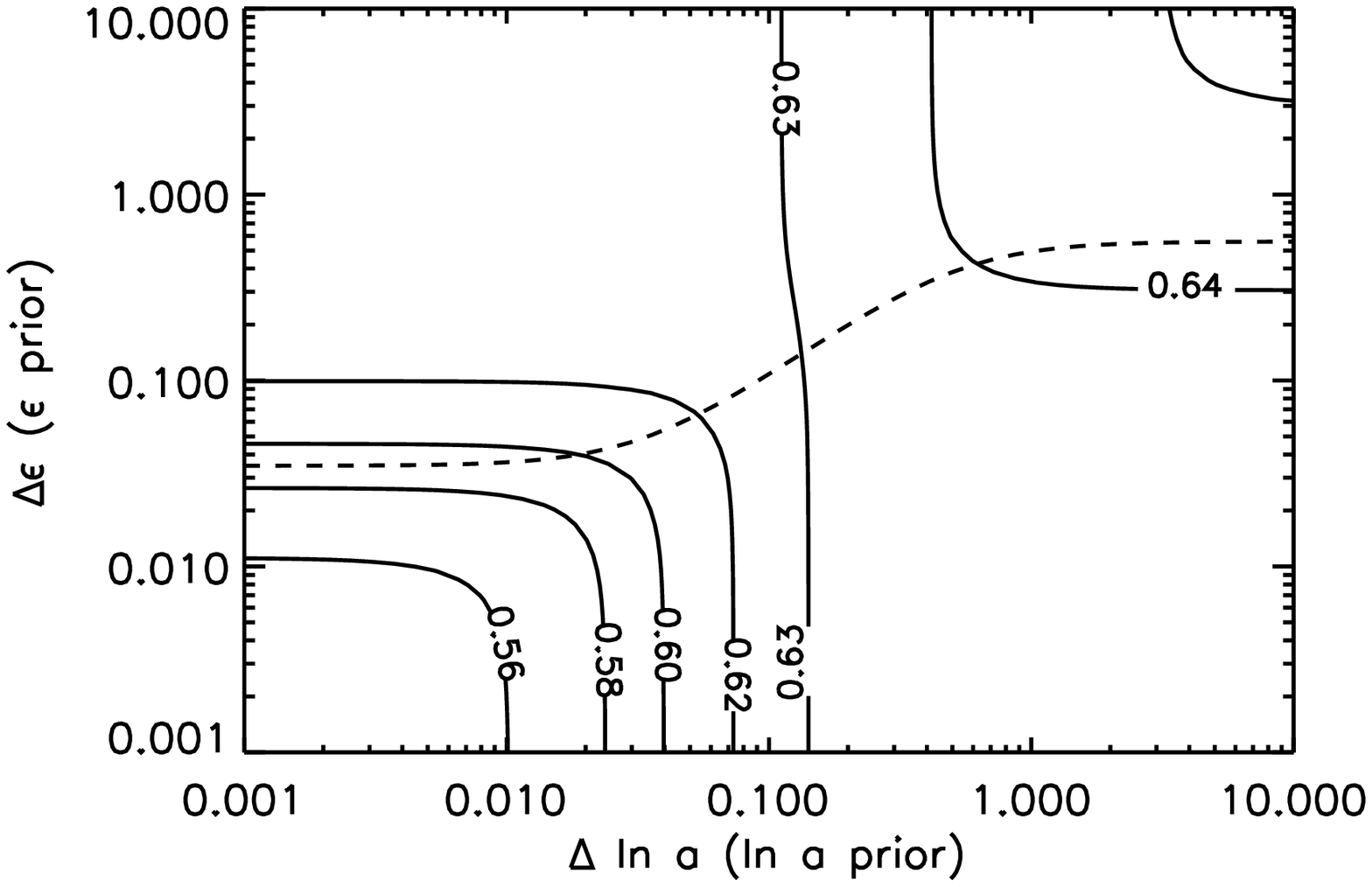}
\caption{{\it Top panel:} Error of a joint shear+density experiment as a function of the priors $\Delta \ln \bar n$ and $\Delta q$. 
The errors have been normalized to those of a standard shear-only analysis.
In these units, perfect knowledge of $\bar n$ and $q$ results in $\sigma_{\ln M}=0.62$, while a complete lack of knowledge
results in $\sigma_{\ln M}=0.94$.  The perfect knowledge limit
is nearly saturated with $\Delta \ln \bar n=0.01$ and $\Delta q=0.1$.  
{\it Bottom panel:} Same as above, but now as a function of the priors $\Delta \ln \bar a$ and $\Delta \epsilon$,
assuming additional priors $\Delta \ln \bar n =0.01$ and $\Delta q=0.1$ (solid curves).  The limiting values are 
$\sigma_{\ln M}=0.56$ when $\bar a$ and $\epsilon$ are fixed, and $\sigma_{\ln M}=0.66$ when neither quantity is known a priori.  
The perfect knowledge limit is nearly saturated at $\Delta \ln \bar a =0.01$ and $\Delta\epsilon =0.01$.  The dashed
curve shows the a-posteriori error in $\epsilon$ as a function of $\Delta \ln \bar a$ when no priors are placed on $\epsilon$,
assuming a source density $\bar n = 10\ \mbox{gals}/\mbox{arcmin}^2$.   Larger source densities will lead to tighter constraints
in $\epsilon$.
}
\label{fig:priors}
\end{figure} 


The top panel in Figure \ref{fig:priors} shows the error of a shear+density analysis as a function of the priors $\Delta \ln \bar n$ and $\Delta q$.
In all cases, the error is normalized to that of a standard shear-only experiment. 
In these units, the minimum error achieved by a joint analysis when both
$\bar n$ and $q$ are perfectly known is $\sigma_{\ln M}=0.62$, 
while complete ignorance of these parameters results in  $\sigma_{\ln M}=0.94$.
A priori knowledge of $\bar n$ and $q$ at the level of $\sigma_{\ln \bar n}=0.01$ and $\sigma_q = 0.1$
is nearly as effective as having perfect knowledge of said quantities.  Such level of a priori knowledge should
be easily achievable.

The solid curves in the bottom panel of Figure \ref{fig:priors} trace the contours of the error in log-mass for our fiducial
cluster in a joint shear+density+size analysis. 
We have assumed realistic priors $\Delta \ln \bar n =0.01$ and 
$\Delta q=0.1$ as per the above discussion, and we used $\ln \bar a$ as a parameter rather than $\bar a$ to enforce
positivity.  With our adopted priors, the error achieved
when $\ln \bar a$ and $\epsilon$ are perfectly known is $\sigma_{\ln M}=0.56$.  To saturate this limit, both $\ln \bar a$ and $\epsilon$ 
must be known at the $\approx 0.01$ level, with the error modestly increasing in the $0.01-0.1$ range.   
Once $\ln \bar a$
and $\epsilon$ are known to $10\%$ level precision, further degradation occurs slowly.
No knowledge of $\bar a$ and $\epsilon$ results in $\sigma_{\ln M}=0.66$.
We expect measuring $\bar a$ should be no more difficult than measuring $\bar n$, and thus $\Delta \ln \bar a =0.01$ should be easily
achieved.  
Saturating the $\epsilon$ bound, on the other hand,
is likely to prove difficult if not impossible, particularly for ground-based imaging.  Thus, in practice we expect to
be able to nearly saturate the limit in which $\epsilon$ is left free, but $\bar n$, $q$, and $\bar a$ are perfectly known
(dashed ellipse in Figure \ref{fig:err_wpriors}).  In this limit, $\sigma_{\ln M}=0.59$, a $40\%$ improvement over the standard
shear-only analysis.

There is an additional benefit to operating in this limit.
The dashed curve in the bottom panel of Figure \ref{fig:priors} shows the forecasted error in $\epsilon$ for a joint density+shear+size analysis
as a function of the prior $\Delta \ln \bar a$ when no prior is placed on $\epsilon$. 
We see that even with the relatively low source density $\bar n =10\ \mbox{sources}/\mbox{arcmin}^2$ we have assumed, 
a joint analysis can recover an unknown bias $\epsilon$ to high precision.  Assuming $\ln \bar a$ is known at the $1\%-10\%$ level,
the corresponding constraint in $\epsilon$ is $\sigma_\epsilon\approx 0.04-0.1$.
Note, too, that unlike the relative uncertainty between a joint lensing experiment and a shear-only analysis, the absolute error on $\epsilon$
is in fact independent on the assumed source density.  In particular, stacked weak lensing analysis, in which the effective source density
can be a hundred or even a thousand times larger, will have a correspondingly tighter constraint on $\epsilon$.
This is an incredibly useful result: 
if $\epsilon$ is believed to be known a priori based on our understanding of PSF subtraction, the joint analysis proposed here allows us to
explicitly check whether or not $\epsilon$ is indeed consistent with zero at the expected level.  This is a highly non-trivial
systematics check on PSF subtraction, particularly when $\ln \bar a$ is known a priori.


\section{Summary and Discussion}
\label{sec:conclusions}

Gravitational lensing changes the density on the sky, size, and ellipticity of background galaxies.  
We have explored the extent to which a joint weak lensing mass calibration
experiment that combines all three signatures 
can improve upon the standard shear-only approach.
To perform this analysis, we must introduce 4 additional nuisance parameters: the mean source density $\bar n$, the lensing parameter
$q$ governing the source density boost due to gravitational lensing, the mean source area $\bar a$, and a PSF subtraction residual $\epsilon$ that
allows for size estimates to be systematically biased. 


\begin{figure}
\epsscale{1.20}
\plotone{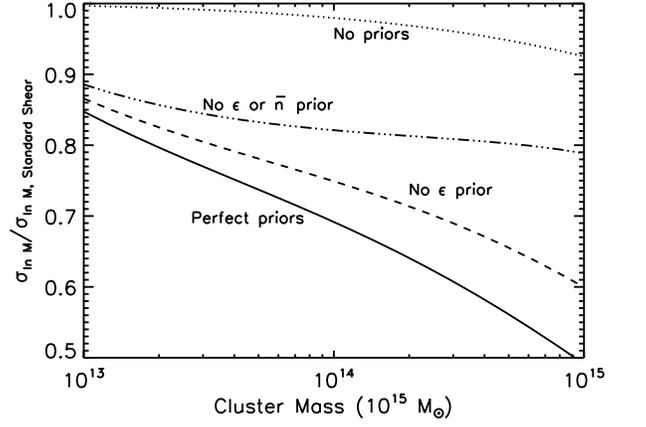}
\caption{Error in log-mass as a function of halo mass for a joint shear+density+size weak lensing experiment, normalized
to the standard shear-only result.  Different curves correspond to different priors of the nuisance parameters $\bar n$ (source
density), $q$ (lensing bias parameter), $\bar a$ (mean source area), and $\epsilon$ (residual PSF correction to source area).
The three cases we consider are no priors on nuisance parameters (dotted), no priors on $\epsilon$ and $\bar n$ (dash-dot),
no priors on $\epsilon$ (dashed), and all nuisance parameters fixed (solid).  We expect real-world experiments will fall
somewhere between the dashed and dash-dot curve.
}
\label{fig:mass}
\end{figure} 


Figure \ref{fig:mass} summarizes our results.
The figure shows the ratio between the uncertainty in log-mass for a joint analysis
relative to the standard shear-only result.  This ratio is plotted as a function of halo mass, and all errors are marginalized
over all other parameters.  The fiducial value for the halo--concentration as a function of mass is taken to be
\be
c(M)=5.0\left (\frac{M}{10^{14}\ \msun} \right)^{-0.1}
\ee
in agreement with the results of \citet{netoetal07}.
The various different curves correspond to different sets of priors: no priors on nuisance parameters (dotted), no priors on $\epsilon$ or $\bar n$ (dash-dot),
no priors on $\epsilon$ (dashed), and all nuisance parameters fixed (solid).

In the limit that no nuisance parameters are known a priori, a joint analysis provides only a marginal improvement
relative to the shear-only case (dotted curve).  In practice, however, we expect the lensing parameters $\bar n$, $q$, and $\bar a$
to be known with sufficient precision to saturate the limit in which these quantities are perfectly known (dashed line). For such a scenario,
the precision of weak lensing mass calibration experiments can improve by as much as $30\%$ relative to the standard shear-only case.
Further improvements are possible if $\epsilon$ can be controlled at the few percent level, which is likely to be a challenging demand,
at least for ground-based surveys.  

Throughout, we have not incorporated source clustering in our error estimates.  While in principle source clustering affects 
all lensing measurements (see the discussion in \cite{paperA}), 
the density measurement is affected directly and most significantly.   Source clustering will be important if 
$\xi(\bar\theta) (\bar n A)^{1/2} \gtrsim 1$, where $\xi(\theta)$ is the source
galaxy angular correlation function, and $\bar\theta = \sqrt{r \Delta r}/d_A(z_L)$ is the mean source separation in an annulus of
physical radius $r$ and width $\Delta r$ at redshift $z_L$.   Moreover,
contamination of the source sample by galaxies associated with the lens, or obscuration of background
galaxies by the optical counterpart of the lens, may lead to systematic error in the density measurement.
A careful characterization of these errors is beyond the scope of this paper, whose purpose is to highlight
that a simultaneous treatment of multiple lensing observables can lead to improved statistical performance
and systematics self-calibration.

However, some insight on the potential systematic impact of source clustering can be obtained by thinking of source clustering as a location-dependent
modulation of the mean source density $\bar n$.  Consequently, removing the prior on $\bar n$ is roughly equivalent to the addition of source
cluster noise.  This scenario is shown in Figure \ref{fig:mass}.  We see that there is an obvious degradation of the mass estimation, but even
in this case a joint analysis is superior to a shear-only analysis.

In light of this discussion, one might expect that the precision that a real-world mass calibration experiment should be able to achieve
lies somewhere between the dashed and dash-dotted line in Figure \ref{fig:mass}.
Interestingly, one by-product of this kind of analysis
is a high precision measurement of $\epsilon$.  The posterior error in $\epsilon$
from a joint analysis with known $\bar n$, $q$, and $\bar a$ is nearly mass independent with $\sigma_\epsilon \approx 0.04$
assuming a source density $\bar n = 10\ \mbox{galaxies}/\mbox{arcmin}^2$.  This empirical determination of $\epsilon$
can be compared to its a priori expectation for a highly non-trivial test of residual PSF subtraction systematics.  Indeed, this is likely
the most significant aspect of these type of analysis: we have explicitly demonstrated that by relying on multiple observables, it is possible
to self-calibrate at least some subset of the parameters characterizing systematic uncertainties.  A much more careful treatment akin to
that of \citet{bernstein09} is required to determine which systematic parameters are best calibrated using these methods.

The degree of systematics-control is likely to be a decisive factor in upcoming surveys, as
the statistical precision of shear-only mass calibration in upcoming photometric surveys
is already expected to reach the $1\%$ level from standard shear-only analysis  \citep{paperA}.
Given that it remains to be seen whether systematics can
be controlled well enough to saturate this limit, analyses which 
allow for some degree of self-calibration are doubly important: not only are the results automatically robust to the
self-calibrated systematics, the empirical determination of systematic parameters can be compared to 
a priori expectations, allowing for crucial consistency tests of our understanding of the experiment.  While we
have not considered any systematics in the shear measurements here, 
we expect that the combination shear+density+size
will be of comparable power in breaking degeneracies when including such systematics.  Indeed, this is exactly the result
of \citet{vallinottoetal10}.
These results provide strong motivation to fully develop this type of joint lensing
analysis as the main weak lensing mass calibration tool for near future data sets such as DES, Pan-Starrs, LSST, and Euclid.

\acknowledgements We would like to thank Bhuvnesh Jain, Matthew Becker, Scott Dodelson, and Alberto Vallinotto for helpful discussions.  
ER is funded by NASA through the Einstein Fellowship Program, grant PF9-00068. 
FS is supported by the Gordon and Betty Moore Foundation at Caltech.

\bibliographystyle{apj}
\bibliography{mybib}

\newcommand\AAA[3]{{A\& A} {\bf #1}, #2 (#3)}
\newcommand\PhysRep[3]{{Physics Reports} {\bf #1}, #2 (#3)}
\newcommand\ApJ[3]{ {ApJ} {\bf #1}, #2 (#3) }
\newcommand\PhysRevD[3]{ {Phys. Rev. D} {\bf #1}, #2 (#3) }
\newcommand\PhysRevLet[3]{ {Physics Review Letters} {\bf #1}, #2 (#3) }
\newcommand\MNRAS[3]{{MNRAS} {\bf #1}, #2 (#3)}
\newcommand\PhysLet[3]{{Physics Letters} {\bf B#1}, #2 (#3)}
\newcommand\AJ[3]{ {AJ} {\bf #1}, #2 (#3) }
\newcommand\aph{astro-ph/}
\newcommand\AREVAA[3]{{Ann. Rev. A.\& A.} {\bf #1}, #2 (#3)}

\end{document}